\documentclass[aps,prl,showpacs,twocolumn]{revtex4}
\usepackage{graphicx}
\begin{document}
\newcommand {\be}{\begin{equation}}
\newcommand {\ee}{\end{equation}}
\newcommand {\bea}{\begin{eqnarray}}
\newcommand {\eea}{\end{eqnarray}}
\newcommand {\nn}{\nonumber}

\title{Geometric fluctuations in a two-dimensional quantum antiferromagnet}
\author{ Anuradha Jagannathan$^a$, Beno\^it Dou\c{c}ot$^b$, Attila Szallas$^c$ and Stefan Wessel$^d$}
\affiliation{$^a$Laboratoire de Physique des Solides, CNRS-UMR 8502, Universit\'e
Paris-Sud, 91405 Orsay, France }
\affiliation{$^b$ LPTHE, Universit\'e Pierre et Marie Curie-Paris 6 and CNRS UMR 7589, Boite 126,
4 Place Jussieu,  75252 Paris Cedex 05}
\affiliation{$^c$ Wigner Research Centre for Physics, Hungarian Academy of Sciences
H-1525 Budapest, P.O.Box 49,  Hungary}
\affiliation{$^d$Institute for Theoretical Solid State Physics,  JARA-FIT,  and JARA-HPC, RWTH Aachen University, Otto-Blumenthal-Strasse 26, D-52056 Aachen, Germany}
\date{\today}

\begin{abstract}
We consider the effects of random fluctuations in the local geometry on the ground state properties of a two-dimensional quantum antiferromagnet. We  analyse the behavior of spins described by the Heisenberg model as a function of what we call \textit{phason disorder}, following a terminology used for aperiodic systems. The calculations were carried out both within linear spin wave theory and using quantum Monte Carlo simulations. An "order by disorder" phenomenon is observed in this model, wherein antiferromagnetism is found to be \textit{enhanced} by phason disorder. The value of the staggered order parameter increases with the number of defects, accompanied by an increase in the ground state energy of the system.

\end{abstract}
\maketitle

Antiferromagnetic Heisenberg models are particularly interesting in the case of two dimensions, where quantum corrections have been studied in a number of lattices, with and without disorder~\cite{rich,manou}. For clean systems, and in the absence of frustrating interactions, the ground state is believed to possess collinear N\'eel semiclassical order even in the extreme quantum limit of spin $S={1}/{2}$. Disordered two dimensional Heisenberg models have also been much studied in the literature. These include models of random exchange, in which the spin exchange couplings are taken from some distribution~\cite{lin,laflor} and site dilution models in which a fraction of sites are unoccupied by spins~\cite{sand,kato,mucc}. In this paper, we consider disorder of geometrical origin, in which random fluctuations cause a local switching of the connections between the sites. In quasicrystals, this type of fluctuation of the local environments is well known by the name of ``phason flip disorder'', and has been proposed as one of the important mechanisms for the formation of such structures based on their entropy \cite{note}. Such defects could in principle exist in regular crystals as well, and in this paper we address the question of their consequences for the quantum antiferromagnetic Heisenberg model. This type of disorder does not affect the classical energy of the system because the overall number of bonds and the strength of the couplings is not affected by the randomness. Quantum fluctuations are modified, however, and we will see that a type of order-by-disorder phenomenon \cite{villain} occurs in this model.

In contrast to the exchange- and site-disorder models, where antiferromagnetism is weakened and ultimately destroyed for sufficiently strong disorder, phason disorder strengthens the order parameter in our model. The ground state energy is increased by defects, and the interaction between defects.

\begin{figure}[t]
\begin{center}
\includegraphics[scale=0.4]{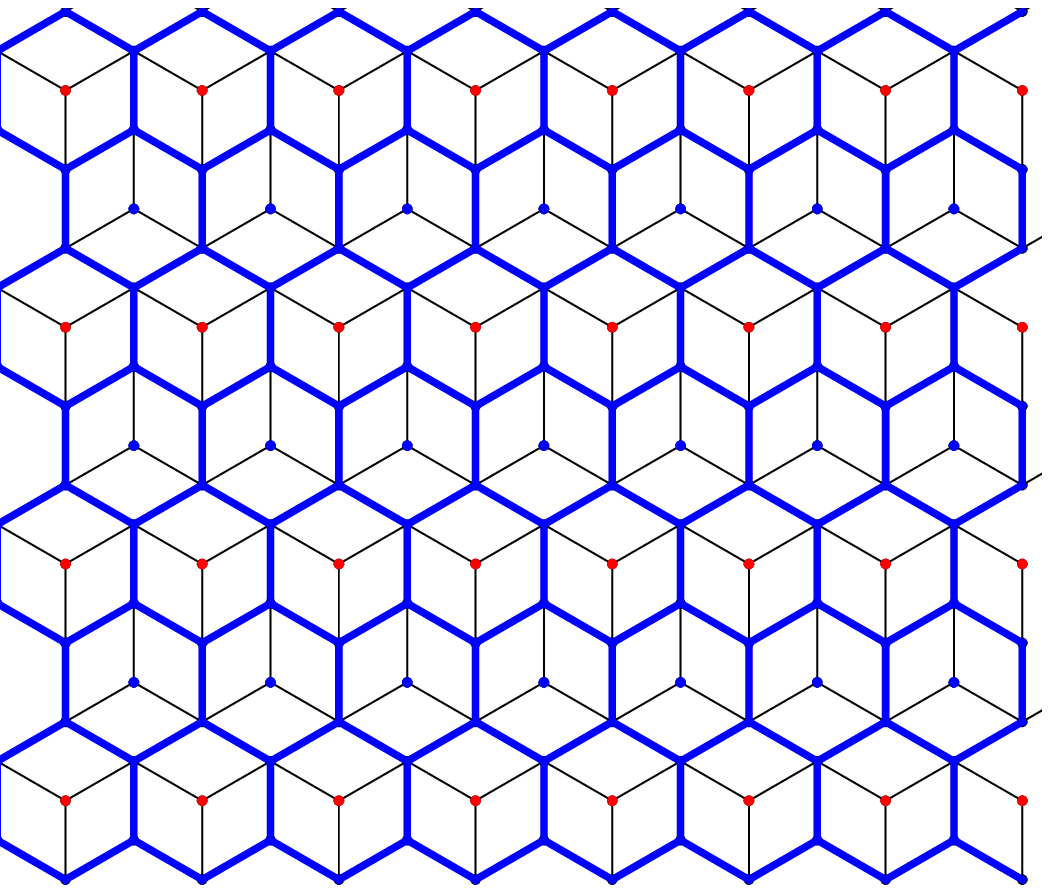}
\includegraphics[scale=0.4]{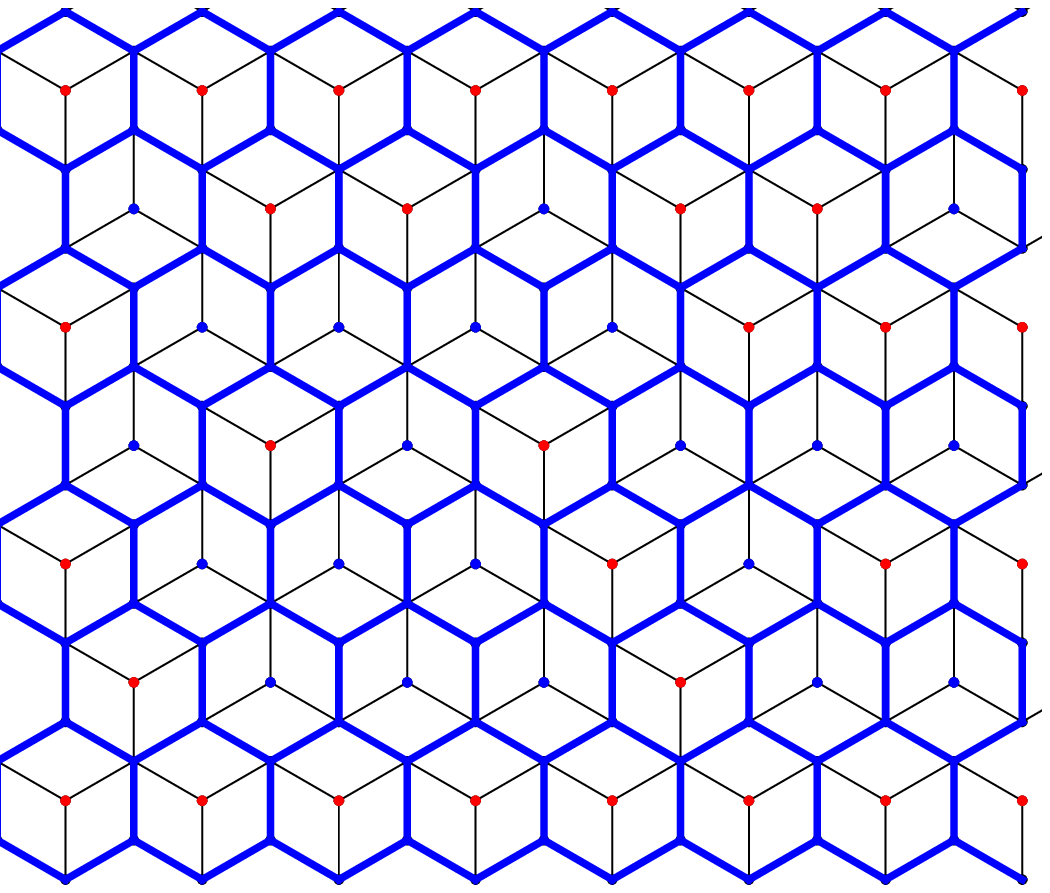}
\caption{ The staggered dice lattice without disorder (left) and with phason disorder (right).}
\label{dicefig.fig}
\end{center}
\end{figure}

\begin{figure}[t]
\begin{center}
\includegraphics[scale=0.4]{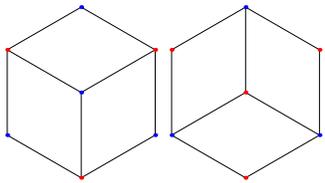}
\caption{Bond configurations before (left, in the U configuration) and after (right, in the D configuration) a single phason flip.}
\label{flip.fig}
\end{center}
\end{figure}

The lattice we consider is the staggered dice lattice, closely related to the usual dice, or T3, lattice. The latter has been investigated for unusual electronic properties associated with its flat band and Dirac cones \cite{berc,moess,t3}. These lattices correspond to different decorations of the honeycomb lattice, as we will now explain. The staggered dice lattice without disorder is shown in the left hand panel of Fig.\ref{dicefig.fig}. The bonds corresponding to the honeycomb lattice are shown with thick lines, while the bonds in the ``interiors" of the hexagons are shown by thin lines. The sites of the hexagonal lattice form hexagonal cages which enclose a site in the interior. The ``interior'' sites are connected to three of the cage sites in one of two possible ways: one can draw three bonds that have the form of a "Y" that faces either upwards (U) or downwards (D). In the ordinary dice (T3) lattice, all hexagons have identical interiors, whereas in the staggered dice lattice, there is an alternation of U and D, along one of the two crystallographic directions. A phason flip, in this system, corresponds to a local fluctuation in which a U-site transforms to a D-site or vice-versa (see Fig.\ref{flip.fig}). When phason flips are randomly introduced in the SDL, one obtains a structure such as the one illustrated in the right hand figure of Fig.\ref{dicefig.fig}. The pure dice lattice (i.e. without disorder) can be realized in optical lattices, as discussed in \cite{berc}, and it would be interesting to similarly realize the phason disordered model experimentally.

In the staggered dice lattice phason flip disorder cannot be increased to arbitrarily high values. In fact, the problem shows a symmetry with respect to the point $\Delta=\frac{1}{2}$, where $0\leq \Delta\leq 1$ is the concentration of phason flips. Linear spin wave theory is used to study the system within the entire range of allowed values, and it gives results in agreement with the results obtained by quantum Monte Carlo calculation using the stochastic series expansion (SSE) method~\cite{sse}.

The dice and staggered dice lattices are both bipartite, so that  nearest-neighbor Heisenberg exchange interactions are unfrustrated. However, there is a significant difference between these models. The dice lattice has two {\it
in-equivalent} sublattices, with a different number of spins on each sublattice $N_A=2N_B$ where A and B denote the two sublattices, and the ground state of the Heisenberg model is thus a ferrimagnet (some properties of which were investigated in Ref.~\onlinecite{jwm}). The ground state of the antiferromagnetic quantum Heisenberg model on the staggered dice lattice of Fig.~\ref{dicefig.fig}a, on the other hand, is an antiferromagnet with two identical sublattices ($N_A=N_B$). The structural and magnetic unit cells are identical, consisting of six sites, two of which correspond to centers of hexagons. The coordination number $z=3$ for the interior sites, while the ''cage" sites have $z=4$ and $z= 5$. The number of sites of each type is equal to $N/3$, and the average coordination number on the staggered dice lattice $\overline{z}=4$.

Consider now the consequences of a phason flip at one of the interior sites that changes the local bond configuration from U to D. In this process, the central site retains the coordination number $z=3$ while the three previous neighbors, belonging to the surrounding hexagon, are replaced by three new ones. As can be seen in Fig.\ref{flip.fig}, the interior site changes its sublattice after such a flip. In order to conserve the equality $N_A=N_B$, we therefore perform the same number of such phason flips on each of the sublattices. When all possible sites are flipped, i.e., all the U-sites are changed to D-sites and vice-versa, the original lattice is recovered upto a global shift along the y-direction (cf. Fig.~\ref{dicefig.fig}a). Thus, the maximum number $N_{max}$ of possible flips $N_{flips}$ is equal to the number of interior sites, $N_{max}=N/3$, where $N$ denotes the total number of lattice sites.  If we define a disorder strength as $\Delta = N_{flips}/N_{max}$, the problem is symmetric with respect to $\Delta =1/2$.

The specific model that we consider is the spin-1/2 quantum antiferromagnetic Heisenberg model, with the Hamiltonian
\bea
H=J \sum_{\langle i,j \rangle} {\mathbf S}_i \cdot {\mathbf S}_j
\eea
where $\bf{S}_i$ are the spin-1/2 operators for the sites $i$ of the staggered dice lattice, and where we assume that the couplings $J>0$ are non-zero for all pairs of spins $i,j$ which are joined by an edge. Since this system is unfrustrated, one expects that the SU(2) symmetry will be broken in the ground state, which will have N\'eel order in which spins are aligned along a direction, say, the $z$-axis.  The local staggered magnetization $m_{si}$ denotes the absolute value of the component of the local magnetic moment in the z-direction (the direction of symmetry breaking), $ m_{si} = \vert\langle S_{iz}\rangle\vert$.

A linearized theory is obtained using the Holstein-Primakoff transformation ~\cite{hols} and keeping only terms quadratic in the bosonic operators of the resulting Hamiltonian.  The ground state energy per site and the local staggered magnetizations were computed numerically for the disordered lattices, as well as for clean samples~\cite{wess2,jag2}. In the latter case one can also reduce the Hamiltonian to a 6 by 6 matrix using the Bloch theorem, and solve analytically at least in the small $\vec{k}$ limit.
The linearized Hamiltonian reads
\bea
\mathcal{H}_{LSW} = -JS(S+1)N_b + JS\mathcal{H}_2, \\ \nonumber
\mathcal{H}_2 = \sum_{\langle i,j \rangle} (a_i^\dag a_i + b_j b_j^\dag +a_i^\dag b_j^\dag + b_j a_i) \label{lswham}
\eea
where the $a_i$ and $b_j$ are the boson destruction operators for sites on the A- and B-sublattice respectively. The number of bonds $N_b=2N$.
After the diagonalization, one obtains

\bea
\mathcal{H}_{LSW} =N E_0 + \sum_\mu\sum_{\sigma} \omega_{\mu\sigma} \gamma_{\mu\sigma}^\dag \gamma_{\mu\sigma} \nonumber \\
E_0=-JS(S+1)N_b/N + JS\sum_{\mu} \omega_{\mu-}
\label{lsw.eqn}
\eea

where the index $\sigma=\pm$ labels the two families of modes that are obtained for this bipartite lattice, and $\mu=1,2,...,\frac{N}{2}-1$ . $E_0$ is the ground state energy per site. The numerical diagonalizations of the LSW Hamiltonian were carried out for several system sizes upto a maximum size of $N=4800$.
For the clean staggered dice lattice, we find that
the ground state energy per site, extrapolated to the thermodynamic limit, gives
$E_0=-0.6517(7)J$  within LSW theory, which compares well to the result  $E_0=-0.6639(1)J$
obtained from QMC simulations.

We now consider the changes due to phason flips.  Consider first the system with two flips, one on each sublattice. The new Hamiltonian can be written $\cal{H'}$=$ \cal{H}$+ $\Delta\cal{H}$, where $\Delta\cal{H}$ is the change due to a flip on the A-sublattice at position $\vec{r}_1$ and one on the B-sublattice at position $\vec{r}_2$. The interaction energy between two phason defects can be calculated to second order in $\Delta\cal{H}$, and a naive power counting argument indicates that the interaction should decay as $1/r^3$ at large distances \cite{notes} (where $r=\vert \vec{r}_1-\vec{r}_2\vert $). Our numerical results for the largest sample studied ($N=4800$) are consistent with a $1/r^3$ decay at large distances, although bigger samples are needed in order to accurately determine the asymptotic behavior. The interaction, which is attractive at short range, tends at large distances to a value $2E_{flip}$, where $E_{flip}\approx 0.057J$ is the energy of creating a single flip. This value can be compared with, for example, the energy of eliminating a single spin, which is about $0.58 J$ in LSW theory \cite{bulut}. It is interesting to compare with results obtained in Ref.~\onlinecite{luesch} for the interaction between two magnetic defects in the square lattice antiferromagnet. The authors carried out a perturbation expansion in $g$, the ratio of impurity spin coupling $J'$ to the pure coupling $J$, and found that the long range interaction between the defects was independent of the coupling and decays as $1/r^3$. The sign of the interaction was found to be negative for defects on opposite sublattices, and positive for defects on the same sublattice. We find similarly that the phason-flip interaction is attractive (repulsive) for phasons on different(same) sublattice. Work on the two flip problem is in progress and will be reported elsewhere.

\begin{figure}[t]
\begin{center}
\includegraphics[scale=0.5]{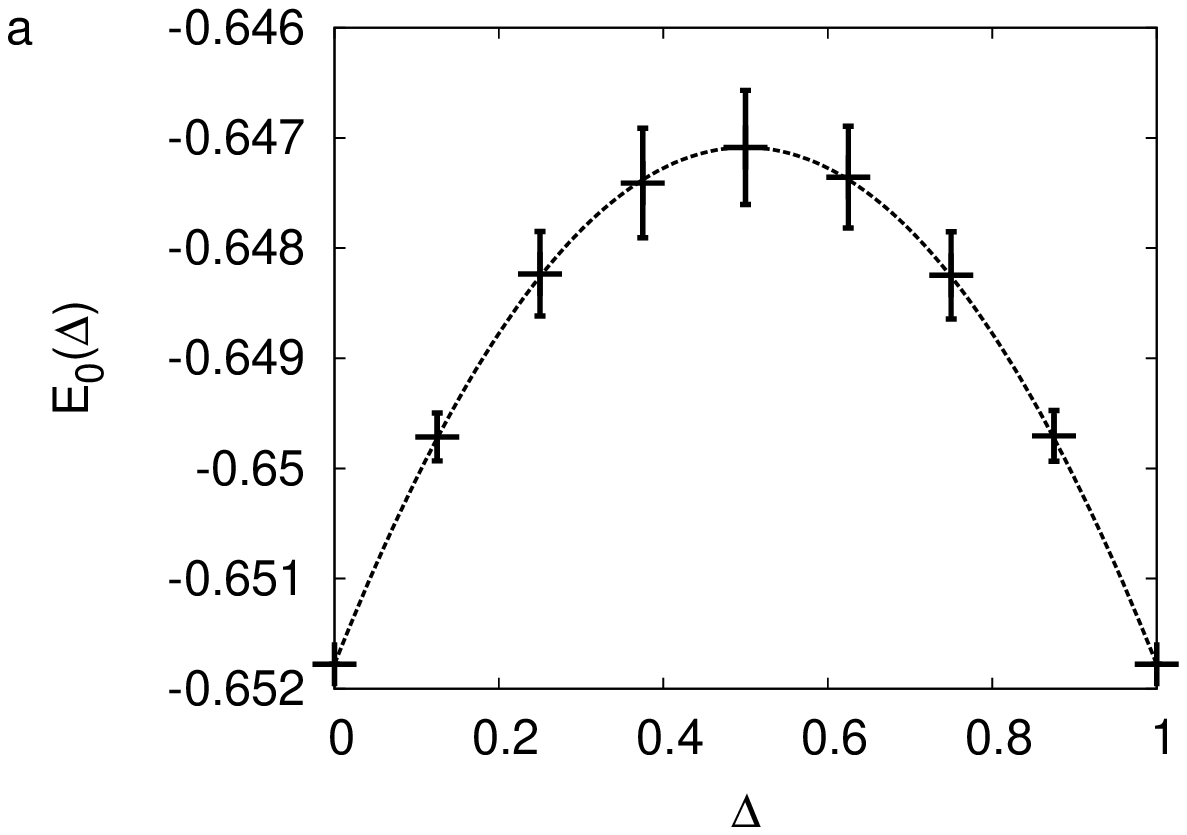}
\includegraphics[scale=0.5]{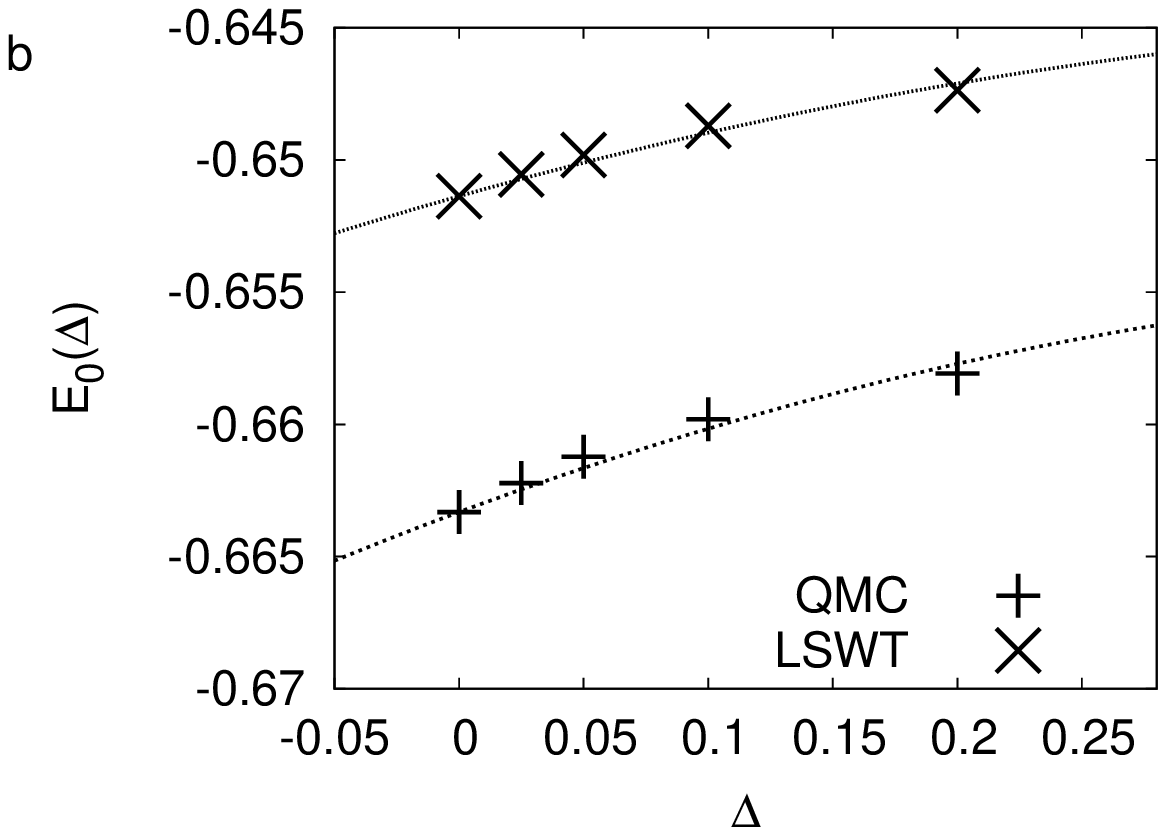}
\caption{
a) The dependence of the ground state energy $E_0(\Delta)$ on the disorder strength $\Delta$ calculated within LSW theory, including a quadratic fit (cf. text). b) Comparison between LSW and QMC results for the range $0<\Delta<0.2$. Here, $N=1200$.}
\label{gse.fig}
\end{center}
\end{figure}

We consider now the evolution of ground state $E_0(\Delta)$ for finite values of the parameter
$0 \leq \Delta \leq 1$. The ground state energy per spin as calculated by LSW is shown in Fig.~\ref{gse.fig}a, while
Fig.~\ref{gse.fig}b shows the comparison between LSW theory and QM in the low-$\Delta$ regime (system size $N=1200$). The ground state energy rises at first linearly, before the curve begins to fall over with increasing
$\Delta$ due to interactions between phason defects. As expected, the curve is symmetric with respect to $\Delta={1}/{2}$.

\begin{figure}[t]
\begin{center}
\includegraphics[scale=0.6]{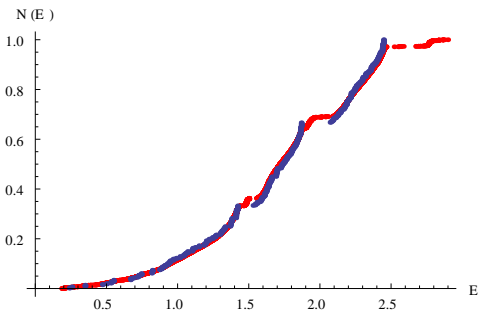}
\caption{(Color online) Plots of the integrated density of states $N(E)$ as a function of energy (in units of $J$) for the pure system (blue) and averaged over the disorder for $\Delta=0.025$ (red) for 1200 spins.}
\label{dos.fig}
\end{center}
\end{figure}

This behavior is well fitted by the expression
\bea
E_0(\Delta) = E_0 + 4(E_{max}-E_0)(\Delta - \Delta^2)
\eea
with $E_0=-0.6514J$ and $E_{max}=-0.6447J$.
The initial linear dependence of $E_0(\Delta)$ is due to the energy of formation of phasons, whose interactions are negligible in the dilute limit. The subsequent curvature of the energy and its quadratic dependence on $\Delta$ can be explained by the fact that the interaction between defects decay very rapidly with distance. This leads to an energy contribution proportional to the square of the number of phason flips. Since the disorder is bounded and there is symmetry around $\Delta=1/2$, this leads to the behavior shown in  Fig.~\ref{gse.fig}. The phason disorder results in \textit{diminishing} quantum fluctuations -- the ground state energy \textit{rises} towards the classical value of $E_{cl}=-0.5 J$.  This behavior is explained by the evolution of the density of states with increasing disorder. The integrated density of states $N(E)$ for a sample of 1200 spins is shown in Fig.\ref{dos.fig}, for the pure system (blue curve) and after averaging over samples for a fixed value $\Delta=0.025$ (red curve). The principal effects of the disorder are to i) fill in the gaps and ii) introduce high energy states around $E\sim 2.75J$ (states arising from the new environments created by disorder, namely sites with z=6). The result is an increase of the integral over energies, which enters in the expression for the LSW ground state energy (Eq.\ref{lsw.eqn}), explaining the observed rise in $E_0$ with increased phason disorder.

\begin{figure}[t]
\begin{center}
\includegraphics[scale=0.5]{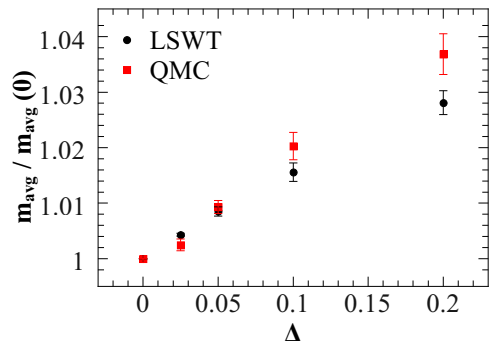}
\caption{Ground state staggered magnetization $\overline{m}_s(\Delta)$ plotted against $\Delta$ in LSW theory and QMC (sample size $N$=1200). }
\label{mags.fig}
\end{center}
\end{figure}

In the clean system, the local staggered magnetization has three different values, with the LSW results of $m_{si} = 0.342, 0.311$ and $ 0.315$ for $z=3,4$ and $5$, respectively. The averaged value determined by LSW is $\overline{m}_s=0.323$, which compares well to the QMC result, $\overline{m}_s=0.3179(5)$ in the thermodynamic limit.  Fig.~\ref{mags.fig} shows the averaged staggered magnetization plotted vs. the disorder strength, as obtained by LSW and by QMC for a sample size of $N=1200$ spins. The averaged value, $\overline{m_s}(\Delta)$ \textit{increases} with $\Delta$ for small $\Delta$. This shows that quantum fluctuations are \textit{reduced} due to the presence of phason flip disorder, in keeping with the ground state energy dependence. This is a type of order-by-disorder phenomenon, and it can be contrasted with the effects of disorder in random bond and site- or bond-diluted antiferromagnets, where quantum fluctuations generically reduce the magnetic order. A possible exception was described in Ref.~\cite{eggert} for a doped quasi-one dimensional compound, although in that system the Neel temperature is observed to fall with doping as well. To explain this manifestation of order-by-disorder a qualitative argument goes as follows: it can be shown by considering a Heisenberg model on a cluster \cite{jwm,jag2} that the local staggered magnetization tends to be largest on sites of smallest coordination number. In effect, transverse spin fluctuations tend to be smaller when the number of nearest neighbors is smaller than the average value. Now, phasons in the pure SDL lead to the appearance of new $z=3$ and $z=6$ sites. The net change in $m_s$ is positive, due to the dominant contribution of small-$z$ sites. This leads to the observed effect, namely an increase of the order parameter with phason disorder. It is interesting to compare the disordered staggered dice lattice with the results obtained in Ref.~\cite{sza3} for a disordered quasiperiodic antiferromagnet. In that case, the ground state energy is lowered -- the quantum fluctuations get bigger as phason disorder is increased. In the quasicrystal, phason flip disorder causes magnon wavefunctions to become more extended and the density of magnons is increased, while in the staggered dice lattice, the situation is reversed.

In conclusion, we have presented quantum effects arising due to geometric disorder in a two dimensional antiferromagnet. The disorder does not introduce frustration, and it conserves the classical ground state energy of the system. Contrarily to other disordered models, phason disorder does not weaken antiferromagnetism which is, on the contrary, strengthened. The changes in ground state energy and the staggered moments indicate that quantum fluctuations $decrease$ with increasing disorder.
This type of disorder could be realized in quasi-two-dimensional structures in which the nearest neighbor environments allow for independent local fluctuations between two different conformations with little or no energy cost. We have discussed in this paper the hexagonal system, in which the two local bond configurations are identical up to a reflection. The notion of phason disorder can be generalized to other systems such as the square lattice, where similar effects are expected to occur (work in progress). Realizations of this disordered model using atoms in optical lattices, along the lines discussed in \cite{berc}, or by Josephson junction arrays can be envisaged in order to experimentally study the phenomena discussed here.

We acknowledge discussions with R. Moessner and thank NIC J\"ulich and
HRLS Stuttgart for allocation of computing time.

\end{document}